# Deep Analog-to-Digital Converter for Wireless Communication

Ashkan Samiee, Yiming Zhou, Tingyi Zhou, and Bahram Jalali
Electrical and Computer Engineering Department, UCLA, Los Angeles, CA 90095

With the advent of the 5G wireless networks, achieving tens of gigabits per second throughputs and low, milliseconds, latency has become a reality. This level of performance will fuel numerous real-time applications, such as autonomy and augmented reality, where the computationally heavy tasks can be performed in the cloud. The increase in the bandwidth along with the use of dense constellations places a significant burden on the speed and accuracy of analog-to-digital converters (ADC). A popular approach to create wideband ADCs is utilizing multiple channels each operating at a lower speed in the time-interleaved fashion. However, an interleaved ADC comes with its own set of challenges. The parallel architecture is very sensitive to the inter-channel mismatch, timing jitter, clock skew between different ADC channels as well as the nonlinearity within individual channels. Consequently, complex post-calibration is required using digital signal processing (DSP) after the ADC. The traditional DSP calibration consumes a significant amount of power and its design requires knowledge of the source and type of errors which are becoming increasingly difficult to predict in nanometer CMOS processes. In this paper, instead of individually targeting each source of error, we utilize a deep learning algorithm to learn the complete and complex ADC behavior and to compensate for it in realtime. We demonstrate this "Deep ADC" technique on an 8G Sample/s 8-channel time-interleaved ADC with the QAM-OFDM modulated data. Simulation results for different QAM symbol constellations and OFDM subcarriers show dramatic improvements of approximately 5 bits in the dynamic range with a concomitant drastic reduction in symbol error rate. We further discuss the hardware implementation including latency, power consumption, memory requirements, and chip area.

## 1. INTRODUCTION

In information theory, the Shannon–Hartley theorem establishes the maximum rate at which information can be transmitted over a communications channel of a specified bandwidth in the presence of noise [1]. The 5G communication systems can achieve tens of gigabits per second throughputs and provide very low latency in tens of milliseconds range. The tremendous increase in the bandwidth challenges traditional signal processing techniques and calls for fresh ideas [2, 3]. Among the critical components in the receiver chain, Analog-to-Digital Converters (ADC) is a key bottleneck because its bandwidth cannot be scaled as easily compared to other electrical components. Sampling rate and effective-number-of-bits (ENOB) are the most important parameters of an ADC [4]. The lower bound on the sampling rate is set by the Nyquist criterion. The ENOB determines the dynamic range of the ADC and must be high enough to handle the dense constellations used in modern communication systems. A popular approach to creating high-resolution wideband ADCs is utilizing multiple channels each operating at a lower speed in the time-interleaved pattern [5, 6, 7, 8]. However, an interleaved ADC comes with its own set of challenges. The parallel architecture is very sensitive to the inter-channel mismatch, timing jitter, clock skew between different ADC channels as well as the nonlinearity within individual channels. Consequently, complex post-calibration is required using digital signal processing (DSP) after the ADC. Traditional methods become challenging for future wideband signals spanning several Gigahertz. One reason is the existence of the parasitic components (resistive, capacitive, and inductive) and their mismatches between interleaved-ADC channels. In narrowband applications, the frequency response of the parasitics is almost constant and a simple scaling can neutralize their effect. However, in wideband systems, the frequency responses can dramatically change with frequency. As a result, new methodologies are required to overcome this issue. Besides, as the feature size of the CMOS technologies is decreasing, traditional CMOS transistor models are not valid anymore and analog circuit designers are having a difficult time designing a robust and accurate circuit. Since traditional DSP methods require knowing the behavior of the ADC circuit, they are less effective in newer CMOS technologies. In contrast to these traditional calibration techniques, neural network (NN) can be used to address the calibration issues of an ADC without requiring knowledge of the circuit behavior. NN has made significant advances in image and voice recognition and synthesis. Recently, several studies [9, 10, 11, 12] have been done about their application in circuit designing and the results have been very promising.

A common NN is made up of several layers connected through multi-dimensional matrix weights. The NN must first be trained with known input/output pairs (ground truth) to find the optimum weights. Different training algorithms such as stochastic gradient descent (SGD) are used to do this [13]. Among network architectures, convolutional networks perform exceptionally well for image processing whereas LSTM networks are best suited for temporal data such as speech processing in which the memory effects are important [14].

Traditional DSP calibration algorithms for ADC errors require a knowledge of the behavior of the error. Owing to a large number of adjustable parameters, neural networks are highly adaptive and can perform ADC calibration without the knowledge of error behavior [15, 16, 17]. In fact, NN has been used for system identification for a long time [18]. Because of these important advantages of the NN over traditional DSP approaches, several studies have been done to prove their ability in removing the non-linear effects in ADCs [19, 20]. However, they utilize primitive neural network architectures and required extensive preprocessing using conventional techniques. They also attempt to calibrate errors for purely analog signals which violate information-theoretic considerations. Consequently, previous attempts have not been successful.

In this paper, we introduce the "Deep ADC" concept where realtime inference by a trained neural network corrects for ADC errors over more than an octave of bandwidth and for a wide range of signal constellations and OFDM subcarrier counts. We show that a single convolutional LSTM neural network that can simultaneously learn and correct the ADC errors including non-linearity induced errors, channel-mismatch, time-skew, phase distortions, and memory effects. We show proof of concept results for an 8G Sample/s 8-channel time-interleaved ADC with the QAM-OFDM modulated data. Dramatic enhancement of approximately 5 bits in the dynamic range is observed with drastic improvement in symbol error rate.

## 2. NEURAL NETWORK ADC CALIBRATION ALGORITHM

The bandwidth (BW) and spectral purity are two of the important topics in cognitive and spectrum sensing. Time interleaving is a widely used method in which multiple similar ADCs quantize the input data series at an aggregate rate that is faster than the operating sample rate of each data converter [5, 6, 7, 8]. Figure 1(a) shows an $M$-channel interleaved ADC. The sampling clock used in each channel is $1/M$ times the final sampling rate and all the channel clocks experience a phase offset of Fs/M from each other. In a time-interleaved ADC, the common sources of error are clock jitter, time-skew, and mismatches between channels (memory effects and non-linearity). These sources of error limit the final dynamic range as measured by SNDR and achievable effective number of bits (ENOB). While traditional DSP algorithms can address these issues in a narrow band system (Hundreds of MHz), they won't be able to calibrate a wideband time-interleaved ADC working at several Gigahertz of Nyquist frequency and providing decent resolution such as 12 bits as used in wireless communication. This is due to mismatches in the frequency responses of channels which is negligible in a narrowband but significant in a wideband scenario.

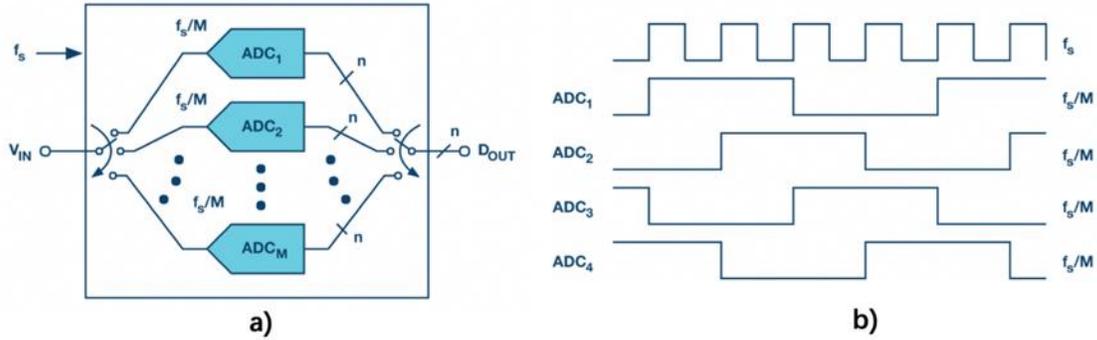

Figure 1. An M-channel interleaved ADC is composed of M individual ADCs working with the rate of F$_s$/M. The sampling clock of each channel has a phase offset of F$_s$/M from the previous channel [5].

To overcome this issue, we propose an advanced NN system that calibrates ADCs over a wide bandwidth. Specifically, we design the demonstrate the system for OFDM/QAM modulated communication signals as this modulation format is the workhorse of wireless communication (LTE, 5G, 6G). The 8-channel Interleaved ADC is modeled in the MATLAB (Figure 2 (a)) and different kinds of error sources are added to make the system behave like a real ADC. Figure 2(b) depicts the block diagram of each ADC channel.

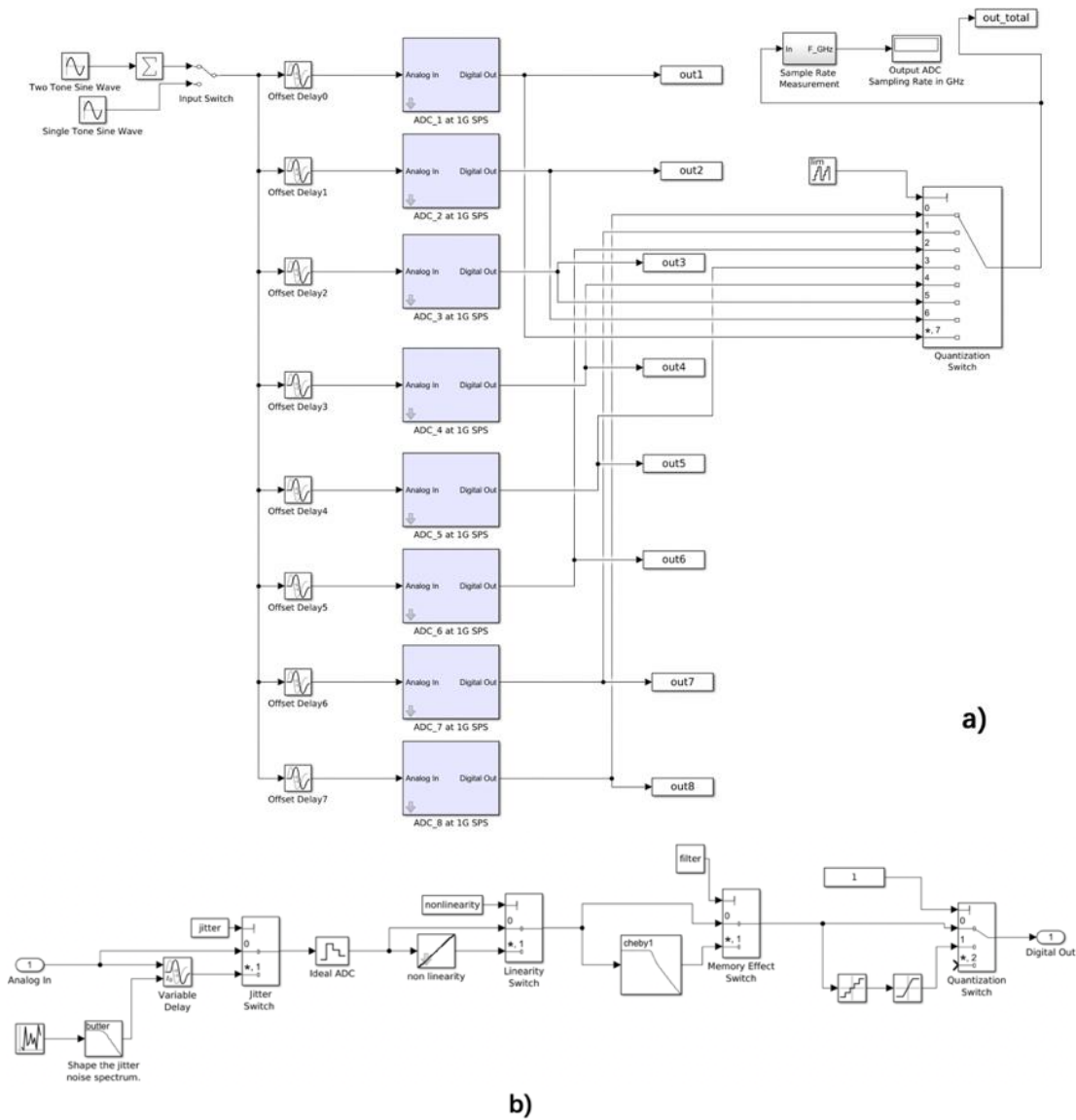

Figure 2. a) Block diagram of an 8-channel interleaved ADC simulated in MATLAB. b) Block diagram of each ADC channel including non-linear effects such as jitter, time-skew, 3$^{rd}$ order nonlinearity, and memory effect.

Each ADC channel includes its own sources of non-linearity such as 3$^{rd}$ order nonlinearity (2$^{nd}$ order is removed by using differential processing), jitter, time-skew, and memory effect (parasitics). The resolution of the ADCs is 13-bit and a full-scale voltage of 1Vp-p. Memory effects include parasitic components that can degrade the performance of the ADC. Figure 3 demonstrates how the NN calibration is trained and implemented in the system design. The network uses current and future samples of the time domain signal. To ensure causality, we buffer the required future time samples and pass the vector to the NN when the buffer is full. Figure 4 shows the architecture of the NN used in this project. It is comprised of a combination of convolutional and LSTM layers. First, several convolutional layers extract the local features temporal features (i.e. with short memory). Then these feature maps are passed through LSTMs layers that capture longer range the memory effects while ensuring causality. Table 1 shows the details of our designed NN architecture.

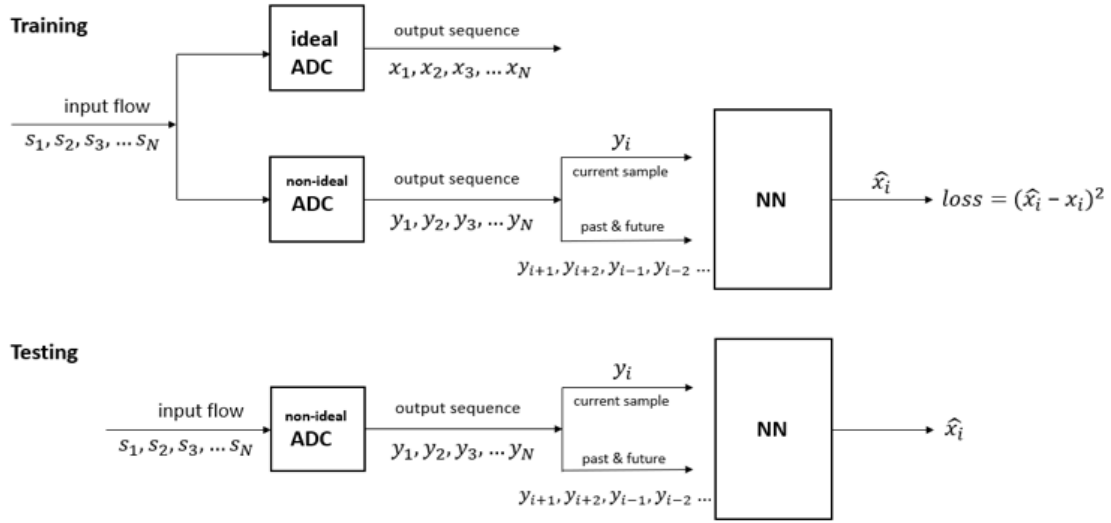

Figure 3. Block diagram of the NN-based ADC error compensation

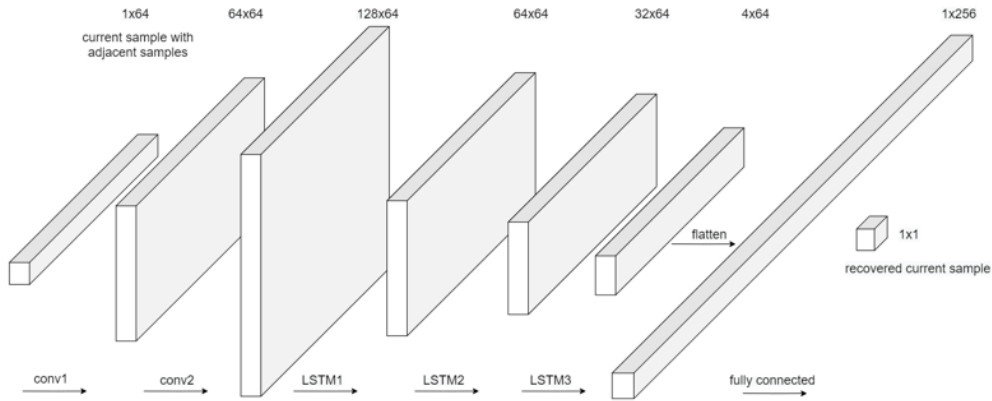

Figure 4. The neural network architecture used for ADC error compensation, the input sequence length (the current sample with adjacent samples) in total is 64 here

| Conv1 | Conv1d (in channel: 1, out channel:64, kernel 3x3, stride:1) + BatchNorm1d + ReLU |
| --- | --- |
|  | Conv1d (in channel:64, out channel:64, kernel 3x3, stride:1) + BatchNorm1d + ReLU |
| Conv2 | Conv1d (in channel:64, out channel:128, kernel 3x3, stride:1) + BatchNorm1d + ReLU |
|  | Conv1d (in channel:128, out channel:128, kernel 3x3, stride:1) + BatchNorm1d + ReLU |
| LSTM1 | LSTM (in features: 128, hidden state features: 64, layer: 1) |
| LSTM2 | LSTM (in features: 64, hidden state features: 32, layer: 1) |
| LSTM3 | LSTM (in features: 32, hidden state features: 4, layer: 1) |
| Linear | Linear (in features: 256, out features: 1) |

Table 1. Details of the neural network architecture for ADC error compensation

## 3. SIMULATION RESULT

In this section, we demonstrate the performance of the Deep ADC and compare its performance under different QAM constellations (64, 128, 256, 512, and 1024). It's important to note that a single network learns the ADC response to all these constellations simultaneously and no retraining is required for inference of different constellations. Only the 256QAM simulation results are presented here. The 64QAM and 1024QAM results are provided in the Appendix section, the rest are similar. The 8-channel interleaved ADC has 13-bit of nominal resolution (12-bit + 1 bit for differential input) and each ADC operates at a sampling rate of 1.024G Sample/s. This provides a total sampling rate of 8.192G Sample/s. The signals under test have a bandwidth of 1GHz with center frequency at 1.3GHz. This signal is generated by passing the QAM symbols through an OFDM modulator with 128 subcarriers and a frequency resolution of 8MHz. Figure 5 shows the peak to average power ratio (PAPR) distribution for each QAM constellation. The time skew for each channel is generated randomly with a uniform distribution. The absolute difference between the maximum and minimum time-skew is 12ps. The jitter has an RMS of 390fs and bandwidth of 20MHz. As it is customary, odd-order nonlinearity is modeled by using the tanh transfer function. Memory effects are simulated by passing each channel through a different Chebyshev analog filter (passband ripple: 1.5 - 6 dB, corner frequency: 5 - 8GHz, order=8) (Figure 6). The corner frequency of the filters is adjusted such that all the signal spectra pass without attenuation but do suffer from amplitude and phase ripples in the passband. The reason for the difference in the frequency responses is to represent the mismatches between channels. This mismatched ripple behavior poses a significant challenge for traditional error calibration algorithms. Figure 7 compares the performance of the NN calibrated ADC with the shifting method for a quantized OFDM symbol in the time domain.

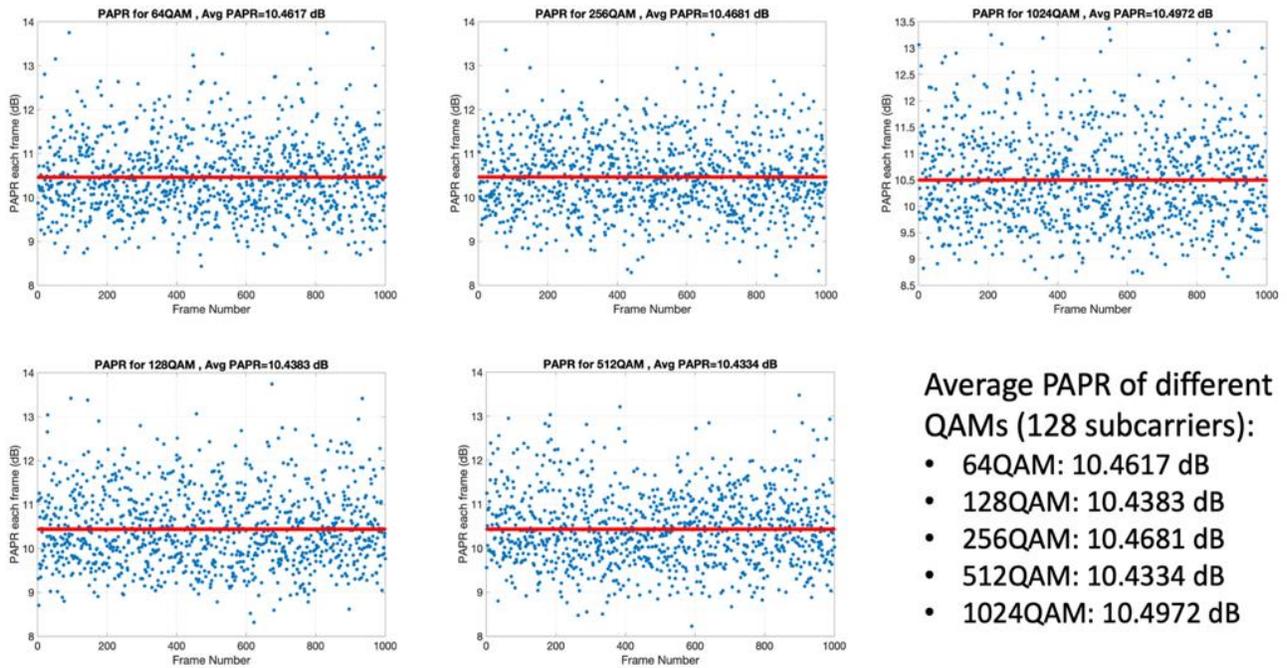

Figure 5. The PAPR distribution for each QAM constellation.

Since the memory effect filters have an almost-linear phase response in the signal spectrum, it would make sense that a shift in time could fix the time domain signal greatly. However, the shifted signal still faces some degradation compared to the ideal signal. Figure 7(c) shows the AI calibrated ADC response. It is interesting to notice that the NN can detect this shift and apply this shift through its network. In Figure 8(a), the time domain error with respect to LSB is shown. Figure 8(b) compares the resulted ENOB by the three approaches. As it is visible from the plots, the NN calibrated method provides a huge improvement compared to other methods. The NN calibrated ADC approximately can improve the ENOB by 8 bits.

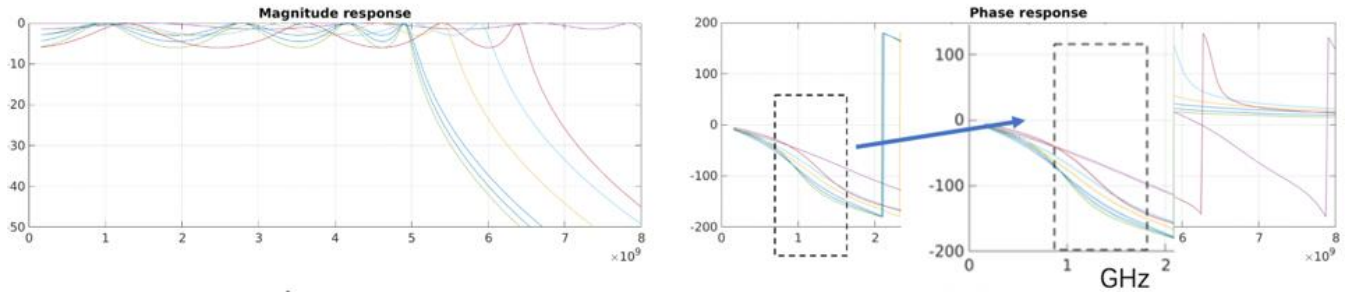

Figure 6. Magnitude and phase response of the filters used to implement memory effects for each of the 8 ADC channels. (passband ripple: 1.5 - 6 dB, corner frequency: 5 - 8GHz, order=8).

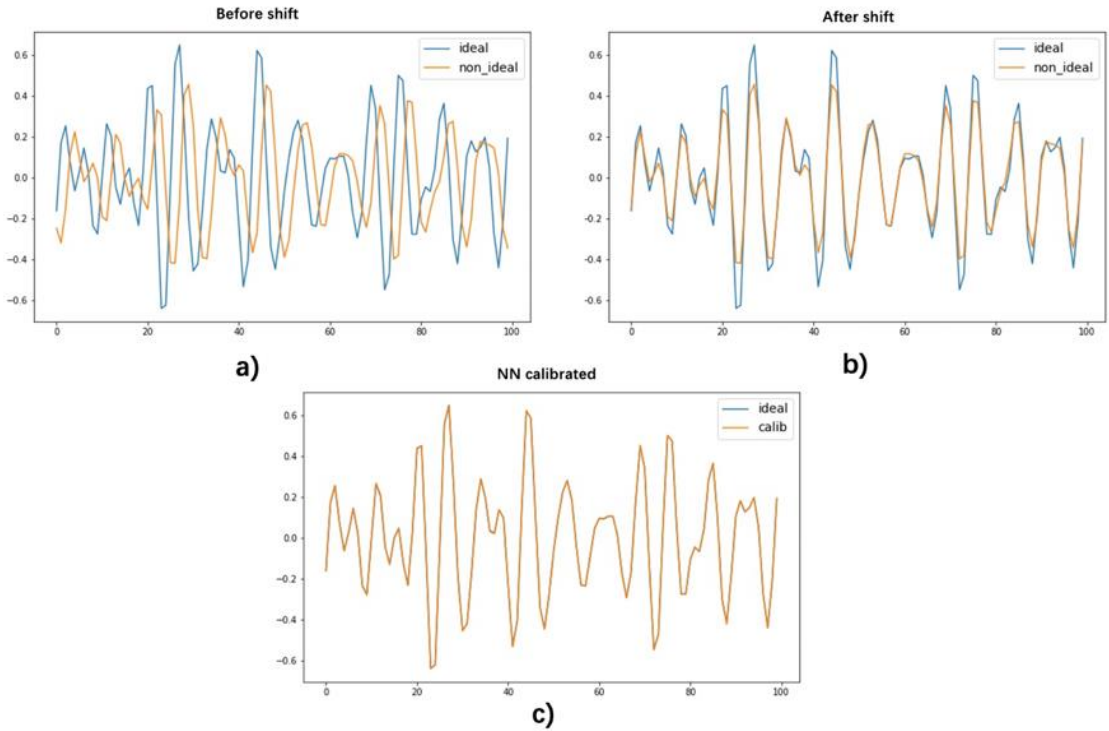

Figure 7. (a) OFDM symbol with ideal (blue) and non-ideal (red) ADC. The data is 256QAM. A large part of the error is a deterministic time shift that can be measured using cross-correlation and easily corrected. This is shown in (b). Figure (c) shows the output of the neural network. Operating as an error compensating postprocessor, the network can learn and correct all the errors including the time shift.

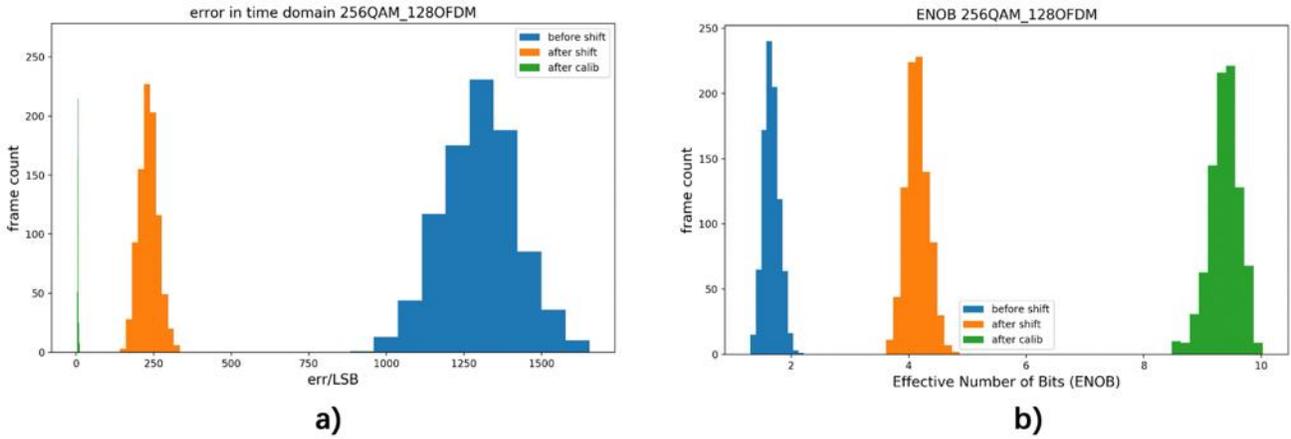

Figure 8. (a) This plot compares the time domain error with normalized to the least significant bit (LSB). The data is 256QAM. (b) Operating as an error compensating postprocessor, the network provides a dramatic improvement in the number of bits.

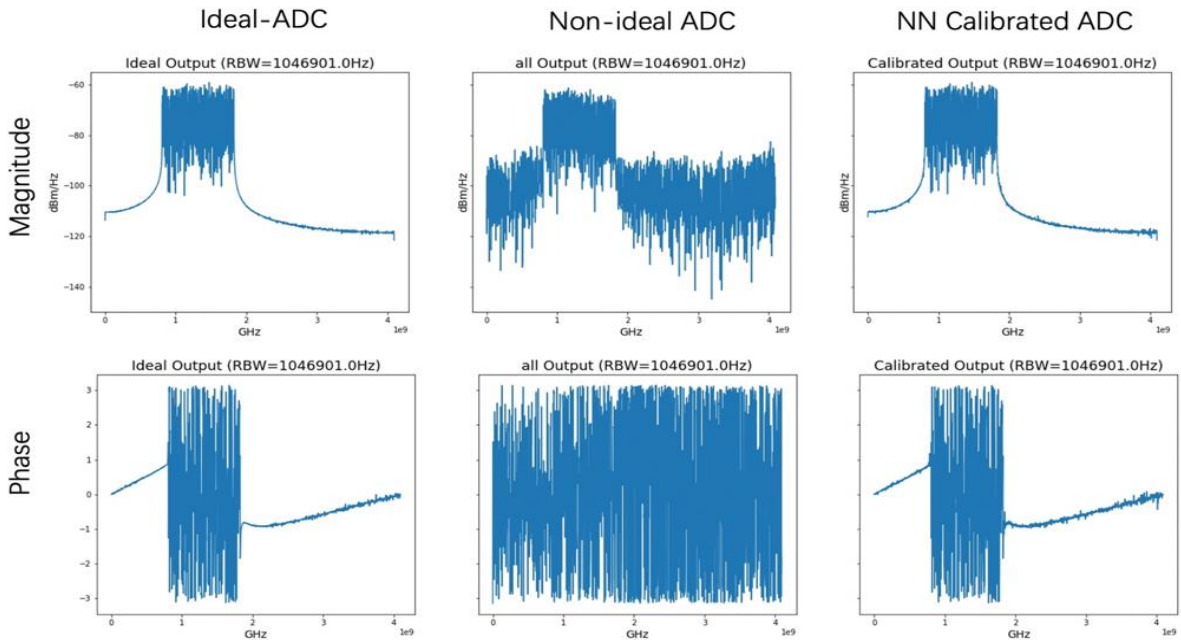

Figure 9. The impact of neural network compensation on the magnitude and phase spectra. Data is 256QAM. From left to right: Frequency spectrum of the one OFDM symbol at the output of an ideal, the non-ideal ADC, and the nonideal ADC after compensation by the neural network. The resolution bandwidth (RBW) is 1MHz.

The simulation results for 64 and 1024 QAM constellations are provided in appendix A and the results for 128QAM and 512QAM are very similar.

## 4. TOWARD HARDWARE IMPLEMENTATION

Although here the training of the NN is done with floating-point calculation, the implementation in either ASIC or FPGA takes place in fixed-point formats. This poses the main challenge for the physical realization of the Deep ADC proposed here. There have been several studies about the quantization of neural networks in both the training and testing phase to realize efficient and accurate fixed-point inference networks [21] [22]. As an example, previous works [23] [24] have demonstrated the FPGA implementation of popular neural network architectures such as convolutional neural networks (CNN) and recurrent neural networks (RNN). Motivated by these successful fixed-point implementations, we believed that the proposed solution can also be implemented in fixed-point formats in hardware. We note that most of the prior works have been classification tasks that are more robust against quantization errors, whereas the ADC calibration is essentially a regression problem that is more sensitive to quantization errors. We anticipate that these quantization errors will place an upper limit on the maximum number of bits that can be achieved.

## 5. CONCLUSIONS

In this project, we presented the concept of Deep ADC where a judiciously designed neural network acts as an error compensating post-processor to enhance the resolution of the ADC. According to the simulation results, this method can calibrate the ADC in a very wide frequency spectrum and improve the resolution for a wide range of OFDM/QAM signals. We show that a single network learns the ADC response to all constellations, simultaneously, and no retraining is required for inference of different constellations. This technology can benefit cognitive and spectrum sensing applications that require wide bandwidth ADCs.

## 6. FUTURE WORK

The main future task is to compress our network by lowering its fixed-point resolution and removing unnecessary weights (pruning). By doing so, the area, memory requirement, and power consumption of the chip are reduced to make it suitable for embedded devices including edge AI.

## REFERENCES

[1] C. Shannon, "A mathematical theory of communication," *The Bell system technical journal,* vol. 27, no. 3, pp. 379-423.

[2] Y. Chiu, B. Nikolic and P. R. Gray, "Scaling of Analog-to-Digital Converters into Ultra-Deep-Submicron CMOS," *Proceedings of the IEEE 2005 Custom Integrated Circuits Conference,* pp. 375-382, 2005.

[3] N. N. Çikan and M. Aksoy, "Analog to Digital Converters Performance Evaluation Using Figure of Merits in Industrial Applications," *European Modelling Symposium,* pp. 205-209, 2016.

[4] R. Van de Plassche, CMOS Integrated Analog-to-Digital and Digital-to-Analog Converters, Springer US, 2003.

[5] G. Manganaro and D. Robertson, "Interleaving ADCs: Unraveling Mysteries," *Analog Dialogue,,* vol. 49, 2015.

[6] B. Brannon, S. Dorn and V. Pai Raikar, "Wideband Receiver for 5G, Instrumentation, and ADEF," Analog Devices.

[7] J. Cao, D. Cui, A. Nazemi, T. He, G. Li, B. Catli, M. Khanpour, K. Hu, T. Ali, H. Zhang and H. Yu, "A transmitter and receiver for 100Gb/s coherent networks with integrated 4× 64GS/s 8b ADCs and DACs in 20nm CMOS," *IEEE International Solid-State Circuits Conference,* pp. 484-485, 2017.

[8] J. Nam, M. Hassanpourghadi, A. Zhang and M. Chen, "A 12-bit 1.6, 3.2, and 6.4 GS/s 4-b/cycle time-interleaved SAR ADC with dual reference shifting and interpolation," *EEE Journal of Solid-State Circuits,* vol. 6, no. 53, pp. 1765-1779, 2018.

[9] K. Hakhamaneshi, N. Werblun, P. Abbeel and V. Stojanovic, "BagNet: Berkeley Analog Generator with Layout Optimizer Boosted with Deep Neural Networks," *IEEE/ACM International Conference on Computer-Aided Design (ICCAD),* pp. 1-8, 2019.

[10] W. Lyu, P. Xue, F. Yang, C. Yan, Z. Hong, X. Zeng and D. Zhou, "An Efficient Bayesian Optimization Approach for Automated Optimization of Analog Circuits," *IEEE Transactions on Circuits and Systems I: Regular Papers,* vol. 65, pp. 1954-1967, 2018.

[11] M. Barros, J. Guilherme and N. Horta, "Analog circuits optimization based on evolutionary computation techniques," p. 136–155.

[12] E. Chang, J. Han, W. Bae, Z. Wang, N. Narevsky, B. Nikolic and E. Alon, "A process-portable framework for generator-based AMS circuit design," *In 2018 IEEE Custom Integrated Circuits Conference (CICC),* pp. 1-8.

[13] L. Bottou, Large-Scale Machine Learning with Stochastic Gradient Descent, Physica-Verlag HD, 2010, pp. 177-186.

[14] H. Palangi, "Deep Sentence Embedding Using Long Short-Term Memory Networks: Analysis and Application to Information Retrieval," *IEEE/ACM Transactions on Audio, Speech, and Language Processing,* vol. 24, pp. 694-707, 2016.

[15] R. P. Lippman, "An introduction to computing with neural nets," *EEE ASSP Mag,* vol. 4, pp. 4-22, 1987.

[16] Y. LeCun, Y. Bengio and G. Hinton, "Deep learning," *Nature,* vol. 2015, no. 521, pp. 436-444.

[17] S. Chu, R. Shoureshi and M. Tenorio, "Neural networks for system identification," *IEEE Control Syst. Mug,* pp. 31-35, 1990.

[18] Z. Fu, W. Xie, X. Han and W. Luo, "Nonlinear Systems Identification and Control Via Dynamic Multitime Scales Neural Networks," *IEEE Transactions on Neural Networks and Learning Systems,* no. 24, pp. 1814-1823, 2013.

[19] A. Baccigalupi, A. Bernieri and C. Liguori, "Error compensation of A/D converters using neural networks," *EEE Transactions on Instrumentation and Measurement,* vol. 45, pp. 640-644, 1996.

[20] S. Xu, X. Zou, B. Ma, J. Chen, L. Yu and W. Zou, "Analog-to-digital conversion revolutionized by deep learning," *Light Sci Appl,* vol. 66, 2019.

[21] R. Krishnamoorthi, "Quantizing deep convolutional networks for efficient inference: A whitepaper.," *arXiv,* 2018.

[22] B. Jacob, S. Kligys, B. Chen, M. Zhu, M. Tang, A. Howard, H. Adam and D. Kalenichenko, "uantization and training of neural networks for efficient integer-arithmetic-only inference," *In Proceedings of the IEEE Conference on Computer Vision and Pattern Recognition,* pp. 2704-2713.

[23] J. Qiu, J. Wang, S. Yao, K. Guo, B. Li, E. Zhou, J. Yu, T. Tang, N. Xu, S. Song and Y. Wang, "Going deeper with embedded fpga platform for convolutional neural network," *In Proceedings of the 2016 ACM/SIGDA International Symposium on Field-Programmable Gate Arrays,* pp. 26-35, 2016.

[24] Y. Guan, Z. Yuan, G. Sun and J. Cong, "FPGA-based accelerator for long short-term memory recurrent neural networks," *Asia and South Pacific Design Automation Conference,* pp. 629-634, 2017.

# Appendix

Below are the test results for the 64 QAM. Please check the simulation section for further information about the test results.

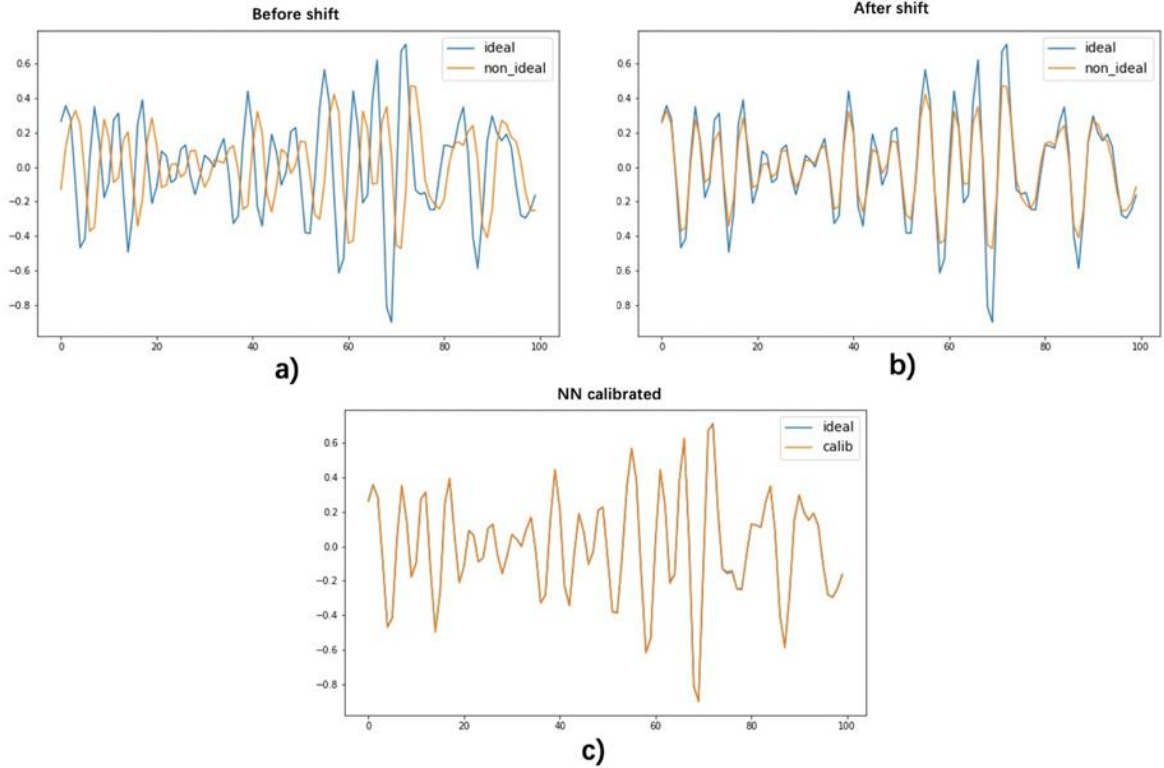

(a) OFDM symbol with ideal (blue) and non-ideal (red) ADC. The data is 64QAM. A large part of the error is a deterministic time shift that can be measured using cross correlation and easily corrected. This is shown in (b). Figure (c) shows the output of the neural network. Operating as an error compensating postprocessor, the network is able to learn and correct all the errors including the time shift.

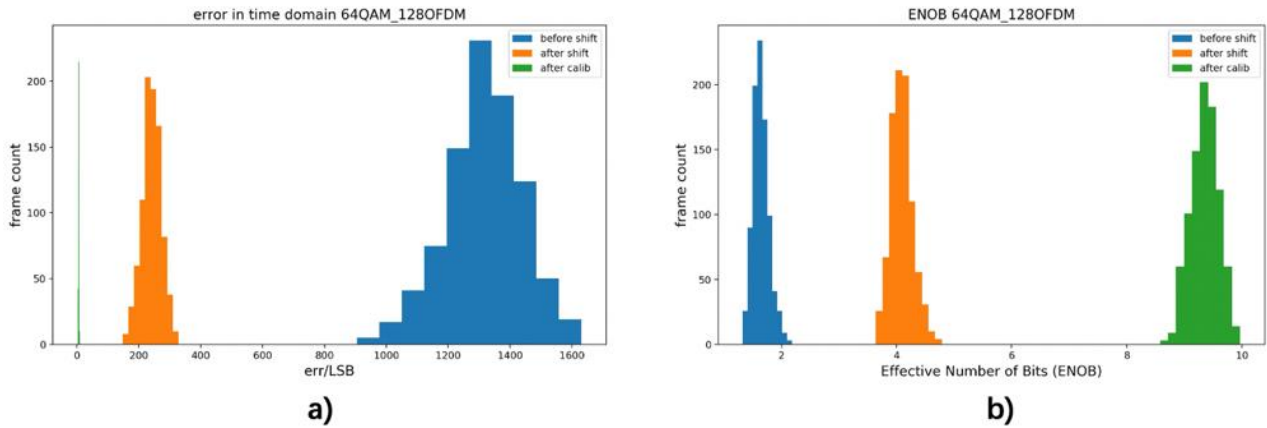

(a) This plot compares the time domain error with normalized to the least significant bit (LSB). The data is 64QAM.
(b) Operating as an error compensating postprocessor, the network provides a dramatic improvement in the number of bits.

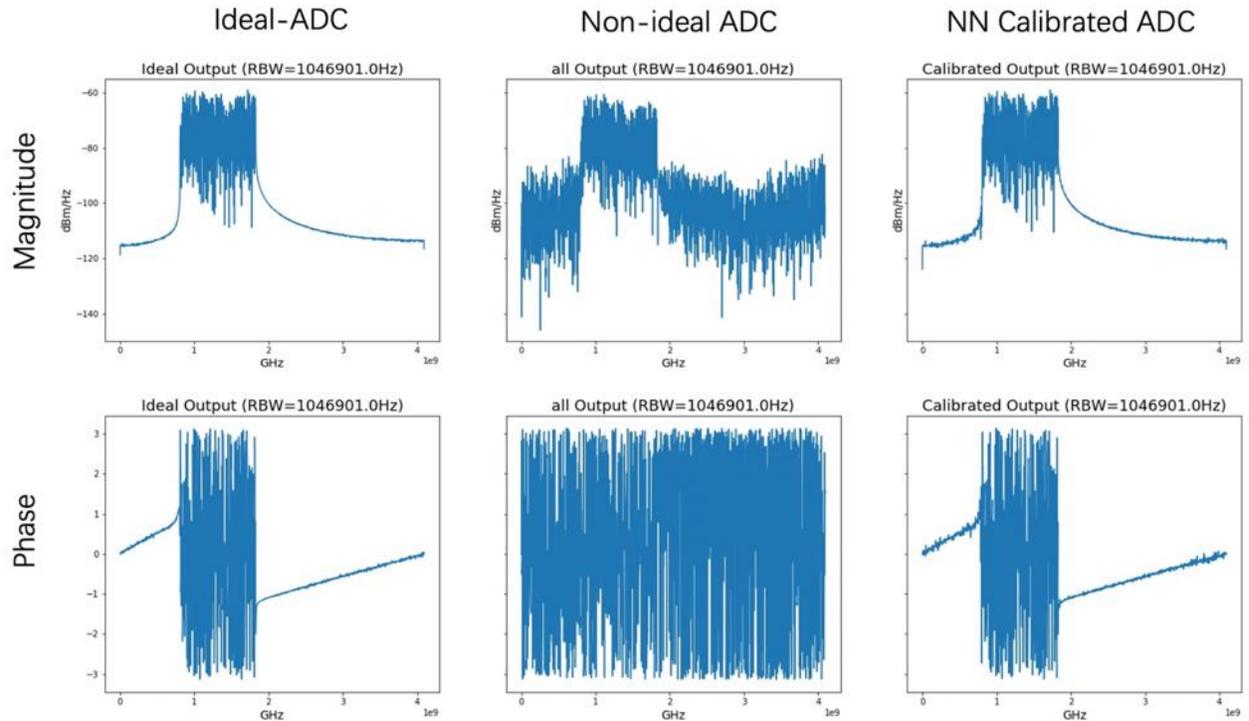

The impact of neural network compensation on the magnitude and phase spectra. Data is 64QAM. From left to right: Frequency spectrum of the one OFDM symbol at the output of an ideal, the non-ideal ADC, and the nonideal ADC after compensation by the neural network. The resolution bandwidth (RBW) is 1MHz.

Below are the test results for the 1024 QAM. Please check the simulation section for further information about the test results.

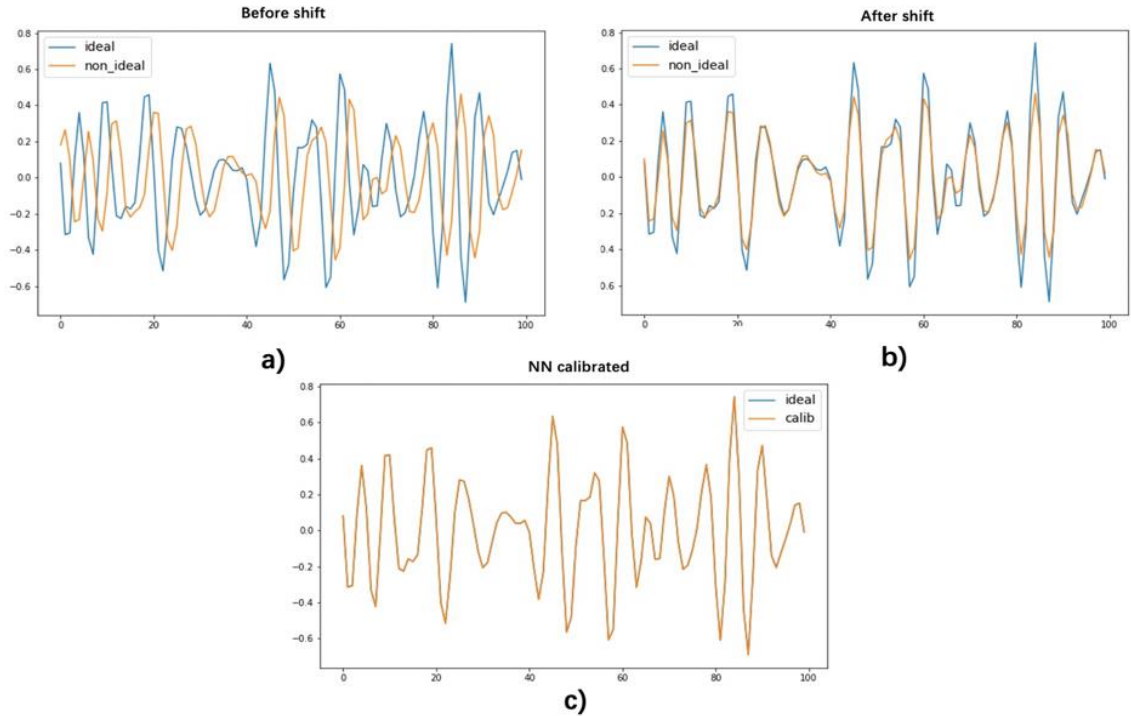

(a) OFDM symbol with ideal (blue) and non-ideal (red) ADC. The data is 1024QAM. A large part of the error is a deterministic time shift that can be measured using cross correlation and easily corrected. This is shown in (b). Figure (c) shows the output of the neural network. Operating as an error compensating postprocessor, the network is able to learn and correct all the errors including the time shift.

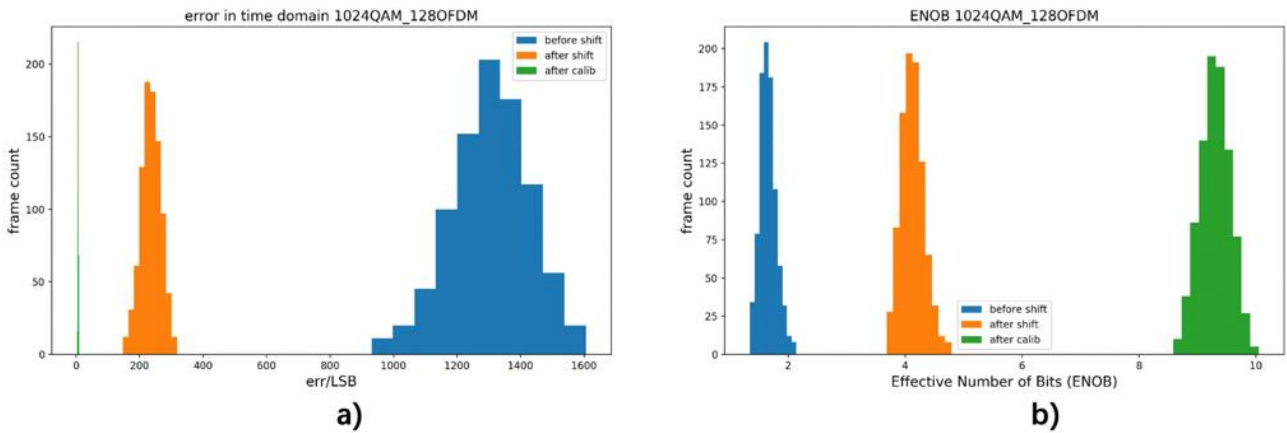

(a) This plot compares the time domain error with normalized to the least significant bit (LSB). The data is 1024QAM. (b) Operating as an error compensating postprocessor, the network provides a dramatic improvement in the number of bits.

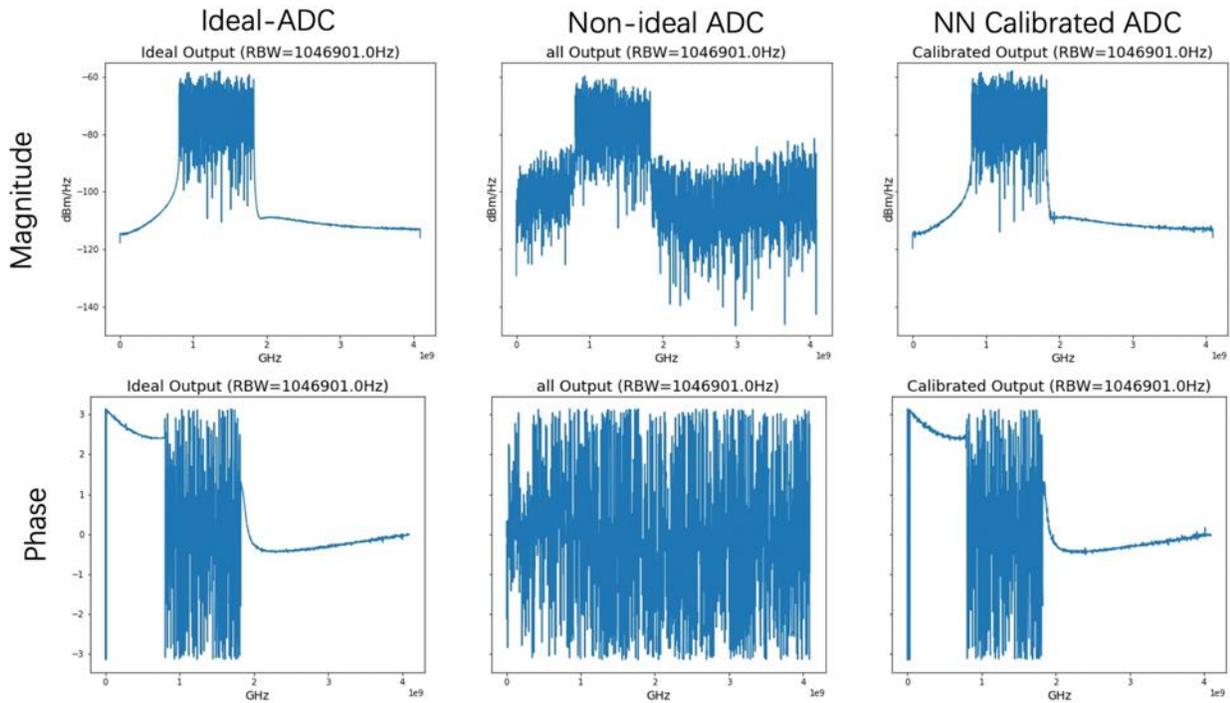

The impact of neural network compensation on the magnitude and phase spectra. Data is 1024QAM. From left to right: Frequency spectrum of the one OFDM symbol at the output of an ideal, the non-ideal ADC, and the nonideal ADC after compensation by the neural network. The resolution bandwidth (RBW) is 1MHz.